\documentclass[aps, pre, twocolumn,groupedaddress]{revtex4} 
\usepackage{graphicx}
\usepackage{amsmath}
\usepackage{amssymb}
\usepackage{ifthen}
\usepackage{booktabs}
\usepackage{gensymb}
\usepackage{graphicx}
\usepackage{dcolumn}
\usepackage{bm}
\usepackage{longtable}

\newcommand{\bb}{\begin{equation}}
\newcommand{\ee}{\end{equation}}
\newcommand{\ba}{\begin{eqnarray*}}
\newcommand{\ea}{\end{eqnarray*}}
\newcommand{\rhor}{\rho({\bf r})}
\newcommand{\dd}{{\rm d}}
\newcommand{\rr}{{\mathbf r}}
\newcommand{\dr}{{\rm d}{\bf r}}

\begin{document}

\title{Symmetry-breaking morphological transitions at chemically nanopatterned walls}

\author{Martin \surname{Posp\'\i\v sil}}
\author{Martin \surname{L\'aska}}
\author{Alexandr \surname{Malijevsk\'y}\footnote{Corresponding author. E-mail address: malijevsky@icpf.cas.cz}}
\affiliation{
{Department of Physical Chemistry, University of Chemical Technology Prague, Praha 6, 166 28, Czech Republic;}\\
 {Department of Molecular and Mesoscopic Modelling, ICPF of the Czech Academy Sciences, Prague 6, 165 02, Czech Republic}}


\begin{abstract}
We study the structure and morphological changes of fluids that are in contact with solid composites formed by alternating and microscopically wide
stripes of two different materials. One type of the stripes interacts with the fluid via long-ranged Lennard-Jones-like potential and tends to be
completely wet, while the other type is purely repulsive and thus tends to be completely dry. We consider closed systems with a fixed number of
particles that allows for stabilization of fluid configurations breaking the lateral symmetry of the wall potential. These include liquid
morphologies corresponding to a sessile drop that is formed by a sequence of bridging transitions that connect neighboring wet regions adsorbed at
the attractive stripes. We study the character of the transitions depending on the wall composition, stripes width, and system size. Using a
(classical) nonlocal density functional theory (DFT), we show that the transitions between different liquid morphologies are typically weakly
first-order but become rounded if the wavelength of the system is lower than a certain critical value $L_c$. We also argue that in the thermodynamic
limit, i.e., for macroscopically large systems, the wall becomes wet via an infinite sequence of first-order bridging transitions that are, however,
getting rapidly weaker and weaker and eventually become indistinguishable from a continuous process as the size of the bridging drop increases.
Finally, we construct the global phase diagram and study the density dependence of the contact angle of the bridging drops using DFT density profiles
and a simple macroscopic theory.
\end{abstract}




\maketitle

\section{Introduction}

The presence of a (flat) solid substrate  can induce local condensation of an ambient gas into a high-density, liquidlike film, the width of which
depends on microscopic interactions and thermodynamic conditions. On a macroscopic level, this ability can be characterized by a contact angle
$\theta$ of a sessile liquid drop deposited on the wall as given by Young's law\cite{rowlinson}
 \bb
  \cos\theta=\frac{\gamma_{wg}-\gamma_{wl}}{\gamma}\,, \label{young}
 \ee
balancing the surface tensions between the wall-gas ($\gamma_{wg}$), wall-liquid ($\gamma_{wl}$) and liquid-gas ($\gamma$) interfaces, respectively.
If $\theta=0$, the droplet spreads over the surface of the wall meaning that in the grand-canonical ensemble the height of the induced liquid film
tends, in an absence of gravity, to diverge and the wall is called to be completely wet. If $\theta>0$, the affinity of the wall towards the liquid
is weaker and the height of the liquid film is only of a microscopic (molecular-scale) dimension; the wall is then partially wet. If $\theta>\pi/2$,
the substrate prefers a contact with the gas, and in an extreme case of $\theta=\pi$ the wall is called to be completely dry.

The wall, embedded into a three-dimensional reservoir of a (simple) fluid, also disrupts the isotropic character of the bulk system. If the wall is
flat and chemically homogenous, the system looses its symmetry in the direction normal to the wall but still preserves translation invariance in the
remaining two Cartesian dimensions tangent to the wall. However, if the (planar) wall is chemically heterogenous, i.e. consist of two or more
species, at least on of these continuous symmetries are broken and so is the shape of the adsorbed liquid film. The generalization of Young's law for
chemically heterogenous walls leads to Cassie's law \cite{cassie1948, quere2003} for the effective contact angle of the sessile drop
 \bb
 \cos\theta^*=\sum_i\chi_i\cos\theta_i \label{cassie}
 \ee
where the summation is taken over all domains of the wall, each characterized by its fractional area, $\chi_i$, and the corresponding Young's contact
angle $\theta_i$.

Cassie's law is a macroscopic result assuming that the interfacial free energy of a composite wall is a sum of contributions from individual regions,
which themselves behave as if they were of infinite extent.  However, the recent advances in experimental techniques \cite{xia1998, qin2010} enabled
to carry out a number of experiments \cite{gau1999, checco2008, bliznyuk2009, luo2010, schafle2010, xue2011, kooij2012, brasjen2013} that revealed
that the phenomenology of wetting on chemically patterned surfaces of various scales is much more complex than that expressed by Eq.~({\ref{cassie}),
especially when a fine structure of the substrate is taken into account \cite{hejazi2014}. This, in turn, invoked theoretical and computational
efforts for a description of liquid adsorption on chemically heterogenous surfaces  by considering further aspects, additional to the surface tension
arguments, such as the relevance of microscopic forces, packing effects, thermal fluctuations, line tension etc.,  all ignored within the original
description of Cassie. These include molecular based simulations \cite{adao1999, schneemilch2003-1, schneemilch2003-2, wang2003, schneemilch2004,
dietrich2005, chen2007, ritchie2012}, stability analysis of liquid structures at microchannels \cite{gau1999, swain1998, lenz1998, lipowsky2001,
lipowsky2005, brinkmann2002}, exact statistical-mechanical arguments \cite{henderson1999, henderson2000}, various modifications of the effective
Hamiltonian model \cite{bauer1999-1, bauer2000, rascon2001, jakubczyk2007}, as well as studies using density functional theory (DFT) approach
\cite{chmiel1994, frink1999, bauer1999-2, porcheron2006, berim2009, yatsyshin2017, mal2017, yatsyshin2018, mal2019, pos2019}.

In terms of the quantitative validity of Cassie's law, the conclusions based on different approaches are contradictory. For example, Gao {\it et al.}
suggest that the use of Cassie's law cannot be recommended as the contact angle behaviour is determined by interactions of the liquid and the solid
at the three-phase contact line alone and thus the interfacial area within the contact perimeter is irrelevant \cite{gao2007}. On the other hand, the
experimental studies dealing with anisotropic wetting \cite{bliznyuk2009, kooij2012} show that the Cassie's law is rather accurate but only in the
direction along the wetting stripe. Finally, there are arguments \cite{marmur2009,schneemilch2004} that Cassie's equation is applicable provided the
wavelength of the surface heterogeneity is sufficiently large, estimated to about $15$ molecular diameters \cite{schneemilch2004}.

\begin{figure}[]
\centerline{\includegraphics[width=\linewidth]{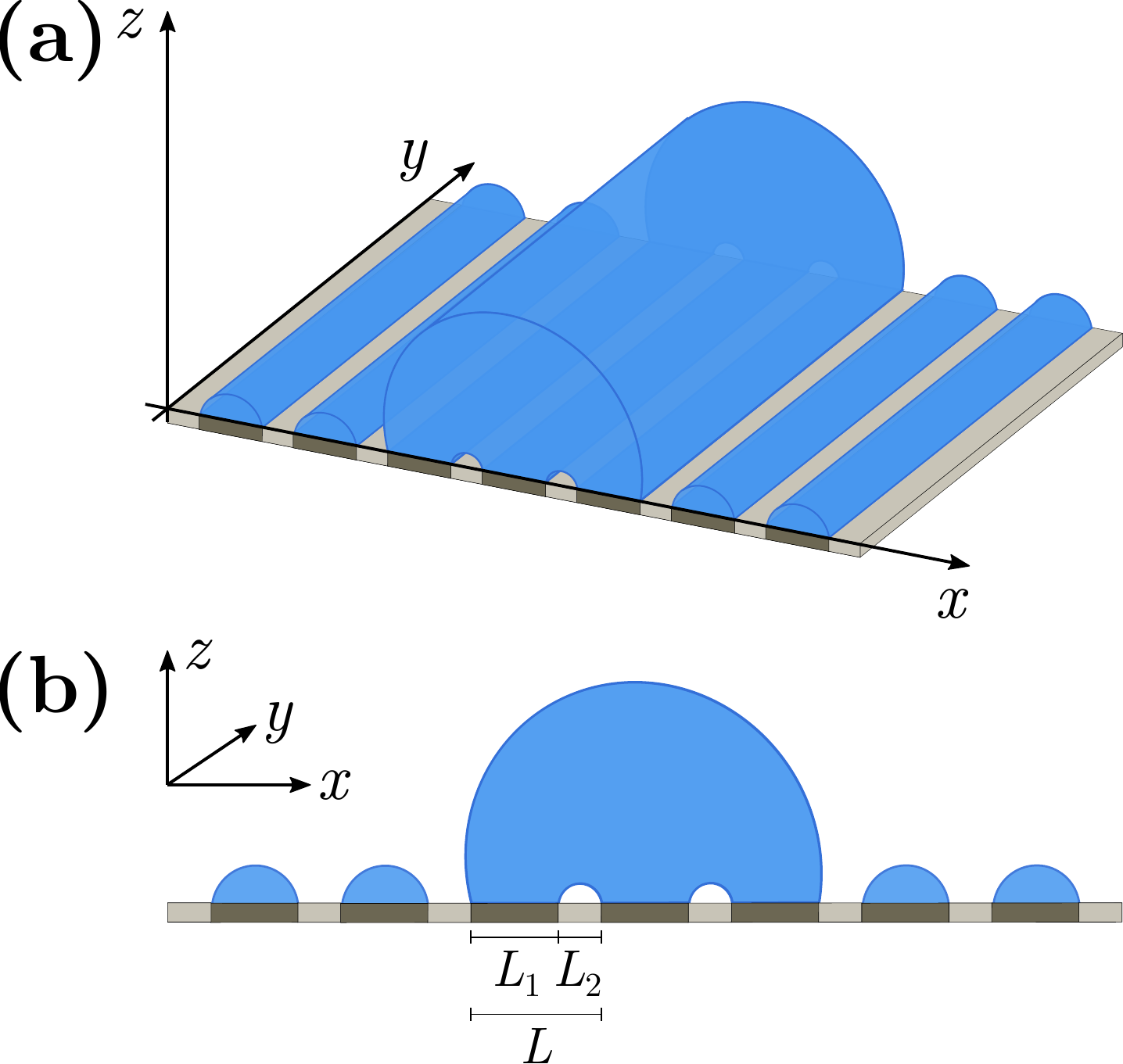}}
 \caption{Sketch of the model substrate consisting of periodically repeating ``hydrophilic''
stripes  (contact angle $\theta=0$) of width $L_1$ and ``hydrophobic'' stripes ($\theta=\pi$) of width $L_2$. The stripes are assumed to be
macroscopically long. The sketch illustrates one of possible liquid morphologies corresponding to a cylindrically shaped drop which connects three
wetting layers; a) top three-dimensional view, b) two-dimensional cross-section.}
  \label{sketch}
\end{figure}

In this work, we present a fully microscopic study of fluid adsorption on nanopatterned planar walls consisting of alternating hydrophilic and
hydrophobic stripes. The parallel stripes are assumed to be microscopically narrow but macroscopically long, so that the system is translation
invariant in the direction along the stripes and the wall potential thus varies only in two dimensions. The external potential exerted by the wall is
periodic but the system itself is finite and closed containing a fixed number of particles.  In contrast to open systems treated in a grand-canonical
ensemble where the fluid structure must follow the symmetry of the external field, it is conceivable for systems with a constrained density to allow
for equilibrium states that break the symmetry of the wall potential, which may bring about a significantly larger number of distinct fluid
configurations. Using a non-local density functional theory (DFT) \cite{evans79} based on Rosenfeld's fundamental measure theory \cite{ros}, the
objectives of this work are as follows: i) We wish to investigate possible fluid morphologies that can be adopted at the given wall, their stability
and the character of the transitions between them as the density of the system varies; ii) we further want to investigate the contact angle
dependence on the system density for droplet liquid morphologies and compare with the macroscopic Cassie's law; iii) finally, we wish to know what is
the impact of the length-scale characterizing the wall patterning on the phase behaviour of the system.

Before concluding the introduction, we underline \emph{a priori} some of the main simplifying assumptions that we have adopted within our analysis
and their expected repercussions. Firstly, we assume that albeit chemically heterogeneous, the wall is perfectly smooth and formed either by
uniformly distributed Lennard-Jones (LJ) atoms (hydrophilic parts) or by atoms interacting with the fluid via only the repulsive bit of the LJ
potential (hydrophobic parts); the net potential of the wall is obtained by summing up the individual LJ potentials over the whole domain of the
wall. Owing to the long-range character of the LJ potential, it follows that the wall attraction is not only exerted exclusively above the attractive
stripes but extends above the repulsive parts as well. This will, presumably, give rise to strong solvation-like forces between liquid drops adsorbed
at the neighboring attractive stripes. As already mentioned above, we assume translation invariance along the stripes which excludes the possibility
of liquid morphologies the shape of which varies in all three dimensions; this includes an important case of cylindrical threads break up, the point
we shall discuss to some detail in the concluding remarks. Although of the finite width, the system is assumed to be a subject of a periodic external
field and the effect of the confining side walls is neglected by imposing periodic boundary conditions along the lateral dimension of the wall. The
system is deemed to be sufficiently high such that the local fluid density does not appreciably change near the top of the box. Furthermore,
throughout the work we neglect the effect of gravity which is justified owing to the typical length-scales (stripes width, atoms diameter,
correlation length) that are all much smaller than the capillary length \cite{degennes}.

\section{Model and density functional theory}

Classical density functional theory is a statistical mechanical tool for a description of structure and thermodynamics of molecular inhomogeneous
systems \cite{ebner76, evans79, evans16}. Within DFT, all the information about the fluid microscopic properties is embraced  in the intrinsic free
energy functional ${\cal{F}}[\rhor]$ of the fluid one-body density (density profile) $\rhor$ and which is typically separated into the ideal gas and
excess contributions
 \bb
 {\cal{F}}[\rhor]={\cal{F}}_{\rm id}[\rhor]+{\cal{F}}_{\rm ex}[\rhor]\,. \label{f_tot}
 \ee
 The ideal-gas part
  \bb
{\cal{F}}_{\rm id}[\rho]=1/\beta\int\dr\rhor\left[\ln(\Lambda^3\rhor)-1\right]\label{ideal}
 \ee
is known exactly, where $\beta=1/k_BT$ is the inverse temperature and $\Lambda$ is the thermal de Broglie wavelength that can be set to unity without
loss of generality.

As the exact knowledge of ${\cal{F}}_{\rm ex}$ is almost never accessible, reasonable and computationally tractable approximations are required. For
atomistic models (as opposed to coarse-grained models), the excess free-energy functional is often treated in a perturbative manner, such that the
excess term is further separated into a purely repulsive hard-sphere and attractive contributions where the latter is usually treated in a mean-field
fashion \cite{sullivan}:
 \bb
 {\cal{F}}_{\rm ex}[\rho]={\cal{F}}_{\rm hs}[\rho]+\frac{1}{2}\int\dr\,\rhor\int\dr'\,\rho(\rr')u_{\rm att}(|\rr-\rr'|)\,.\label{fex}
 \ee
 The attractive tail of the fluid-fluid interaction $u_{\rm att}(r)$ is taken to be a truncated Lennard-Jones-like potential
 \bb
 u_{\rm att}(r)=\left\{\begin{array}{cc}
 0\,;& r<\sigma\,,\\
-4\varepsilon\left(\frac{\sigma}{r}\right)^6\,;& \sigma<r<r_c\,,\\
0\,;&r>r_c\,.
\end{array}\right.
 \ee
which is cut off at the distance of $r_c=2.5\,\sigma$. Here, $\varepsilon$ and $\sigma$ are the interaction parameters that we eventually take as the
energy and length units, respectively. Furthermore, we associate the parameter $\sigma$ with the hard-core diameter of the fluid particles, the
presence of which models repulsive interaction between the fluid atoms and which contributes to the repulsive bit of the excess free energy
(\ref{fex}). This contribution, which is responsible for the short-range correlations between fluid particles and accounts for packing (exclusion
volume) effects, is approximated using the fundamental measure theory (FMT) \cite{ros}
 \bb
{\cal{F}}_{\rm hs}[\rho]=\int \Phi(\{n_\alpha\})\,\dr\,,\label{fhs}
 \ee
where the free-energy density $\Phi$ is a function of six weighted densities $\{n_\alpha\}$ given by a convolution of $\rhor$ with the corresponding
weighting function. Among various versions of FMT \cite{cuesta} we have adopted original Rosenfeld's functional \cite{ros} which provides both a
fully consistent and accurate approach for the fluid state.

In general, the effect of a confining wall is included in DFT as an external field which the wall exerts. In this work, we consider a solid planar
wall consisting of periodically repeating ``hydrophobic'' and ``hydrophilic'' stripes with a periodicity of $L$. The latter are represented by
Lennard-Jones atoms interacting with the fluid particles via the potential
 \bb
   \phi_1(r) = 4\varepsilon_w
  \left[
    \left(\frac{\sigma}{r}\right)^{12} - \left(\frac{\sigma}{r}\right)^6 \label{phi1}
  \right]\,,
 \ee
  the strength of which is controlled by the parameter $\varepsilon_w$.

The ``hydrophobic'' parts of the wall are taken to be purely repulsive by ``turning off'' the attractive contribution to the Lennard-Jones potential
of the wall atoms leading to
 \bb
   \phi_2(r) = 4\varepsilon_w
    \left(\frac{\sigma}{r}\right)^{12}.
 \ee

The net potential of the wall is obtained by integrating the wall-fluid pair potential over the whole domain of the wall. Here we assume that the
stripes are macroscopically long, so that the system can be treated as translation invariant along the stripes (Cartesian axis $y$) and that the wall
atoms are distributed uniformly with a density $\rho_w$. Hence, the wall potential can be written as
 \begin{eqnarray}
  V(x,z)&=&\frac{4}{45}\pi\varepsilon_w\rho_w\sigma^3\left(\frac{\sigma}{z}\right)^9\nonumber\\
  &&+\sum_{n=-\infty}^{\infty} V_{L_1}(x+nL,z)\,. \label{v1}
 \end{eqnarray}
The first term is the total repulsion of the wall, while the second part is a sum over the attractive wall-fluid interaction due to a single
hydrophilic stripe of width $L_1$ represented by the potential $V_{L_1}(x,z)$ which can be expressed as
 \begin{eqnarray}
V_{L_1}(x,z)&=&-4\varepsilon_w\sigma^6\rho_w\int_{x-L_1}^x\dd x'\int_{-\infty}^\infty\dd y'\int_{z}^{\infty}\dd z'\nonumber\\
 &&\times\frac{1}{(x'^2+y'^2+z'^2)^3}\label{v2}\\
  &=&\alpha_w\left[\frac{1}{(x-L_1)^3}-\frac{1}{x^3}+\psi_6(x-L_1,z)-\psi_6(x,z)\right]\nonumber
\end{eqnarray}
where
 \bb
\alpha_w=-\frac{1}{3}\pi\varepsilon_w\sigma^6\rho_w
 \ee
 and
 \bb
\psi_6(x,z)=-{\frac {2\,{x}^{4}+{x}^{2}{z}^{2}+2\,{z}^{4}}{2{z}^{3}{x}^{3} \sqrt {{x}^{2}+{z}^{2}}}}\,.
 \ee

Within the standard formulation of DFT one deals with an open system which is at contact with a bulk reservoir of temperature $T$ and chemical
potential $\mu$. Here, we deal with closed systems with the fixed average fluid density $\rho_{\rm av}$, rather than $\mu$, defined as
 \bb
 \rho_{\rm av}=\frac{\int_V\dr\,\rhor}{V}\,,
 \ee
where $V=L_xL_yL_z$ is the volume and $L_i$ are the Cartesian dimensions of the rectangular system. The equilibrium density profile is obtained by
minimizing the total free energy functional
 \bb
  F[\rhor]={\cal{F}}[\rhor]+\int\dr\rhor V(\rr) \label{frho}
 \ee
where $V(\rr)=V(x,z)$ is the external potential induced by the  wall as given by Eqs.~(\ref{v1}) and (\ref{v2}). The minimization is subject to the
constraint
 \bb
 \int_V\dr\rhor=N\,,
 \ee
 which after substitution of (\ref{f_tot}) and (\ref{ideal}) into (\ref{frho}) leads to the Euler-Lagrange equation
  \bb
 \frac{\rho(x,z)}{L_xL_z}=\frac{\rho_{\rm av}\,{\rm exp}[c^{(1)}(x,z)]}{\int_0^{L_x}\dd x\int_0^{L_z}\dd z\, {\rm exp}[c^{(1)}(x,z)]}\,, \label{el}
  \ee
  where
  \bb
c^{(1)}(x,z)=-\frac{\beta\delta F[\rhor]}{\delta\rhor} \label{c1}
  \ee
is the single-particle direct correlation function \cite{evans79}.

Equation (\ref{el}) is solved numerically on a two-dimensional rectangular grid with an equidistant spacing of $0.1\,\sigma$ using Picard's
iteration. At the $i$-th step of the iteration the new density profile $\rho^{(i+1)}(x,z)$ is obtained from the previous one $\rho^{(i)}(x,z)$ by
computing the one-body direct correlation function (\ref{c1}) following a methodology described in Ref.~\cite{mal13}. Unless stated otherwise, the
dimensions of the system are set to $L_x=L_z=60\,\sigma$ and we impose the periodic boundary conditions along the $x$-axis $\rho(x,z)=\rho(x+L_x,z)$.
The interaction wall parameter is set to $\varepsilon_w=\varepsilon$ for which the wetting temperature of the hydrophilic parts is $T_w\approx
0.8\,T_c$ where $k_B T_c=1.41 \varepsilon$ corresponds to the critical temperature of the bulk liquid-vapour transition. Throughout this work we
consider the temperature $T=0.92\,T_c>T_w$ which means that the hydrophilic parts of the wall tend to be completely wet ($\theta=0$), in contrast to
the rest of the wall which, regardless of the temperature, tends to be completely dry ($\theta=\pi$). At this temperature, the bulk fluid phases
coexist at the chemical potential $\mu_{\rm sat}=3.97\,\varepsilon$ and the corresponding fluid particle densities are $\rho_g=0.10\,\sigma^{-3}$ and
$\rho_l=0.44\,\sigma^{-3}$.

\section{Results}

\subsection{Neutral substrate, $\mathbf{\chi=0.5}$}

\onecolumngrid

\begin{figure}[tb]
\centerline{\includegraphics[width=\linewidth]{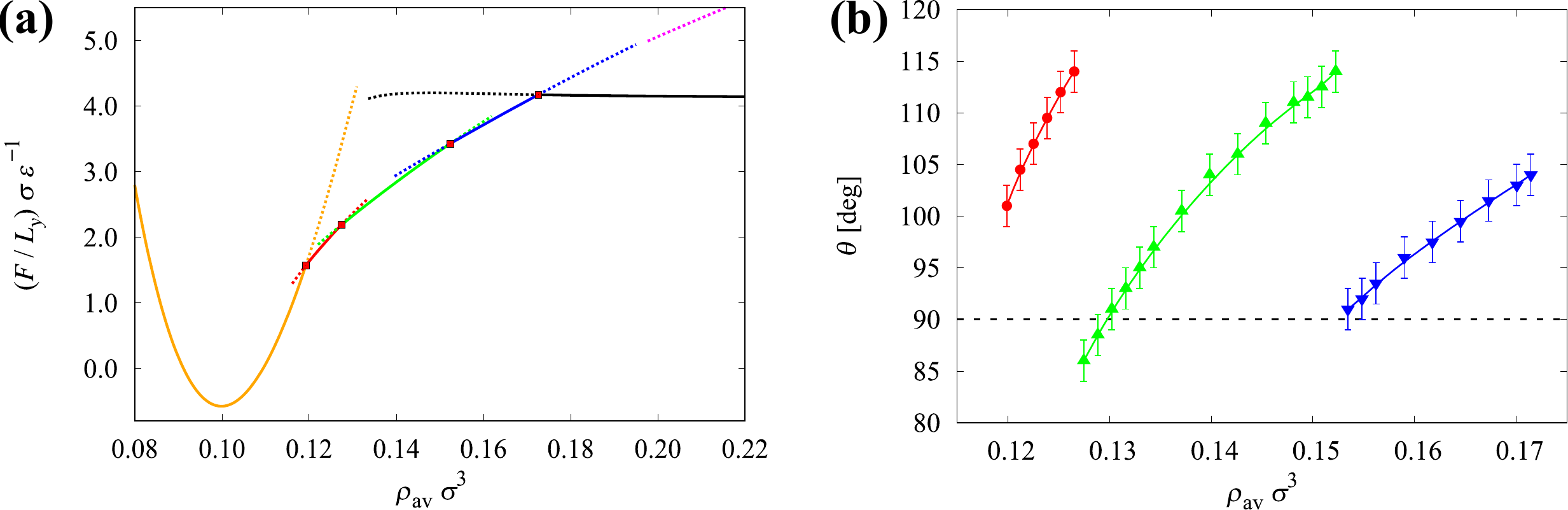}}
 \caption{a) Density dependence of the free energy per unit length for the neutral substrate
($\chi=0.5$) with the stripe widths of $L_1=L_2=5\,\sigma$. The full lines represent the free-energy envelope corresponding to the equilibrium states
of various liquid structures that are separated by first-order morphological transitions highlighted by filled squares. The configuration
corresponding to a microscopic coverage of separated wetting layers (orange curve) is followed by symmetry-breaking states with a drop attached on
two (red), three (green), four (blue) and five (magenta) stripes; the free energy branch of a liquid slab morphology is in black. Also shown are the
metastable extensions of the equilibrium states that correspond to local minima of the free energy functional (\ref{frho}) depicted by dashed lines.
b) Contact angle density dependence of the bridging drop for the corresponding substrates  as obtained from the DFT density analysis. The results
correspond to stable configurations of a drop attached on two (red circles), three (green upward triangles) and four (blue downward triangles)
attractive stripes. The symbols are connected by lines to guide the eye and vertical lines represent estimated inaccuracy in the determination of the
contact angle. The dashed horizontal line shows the value of the Cassie's apparent contact angle as given by Eq.~(\ref{cassie}).} \label{pd5}
\end{figure}

\twocolumngrid


\begin{figure}[]
\centerline{\includegraphics[width=\linewidth]{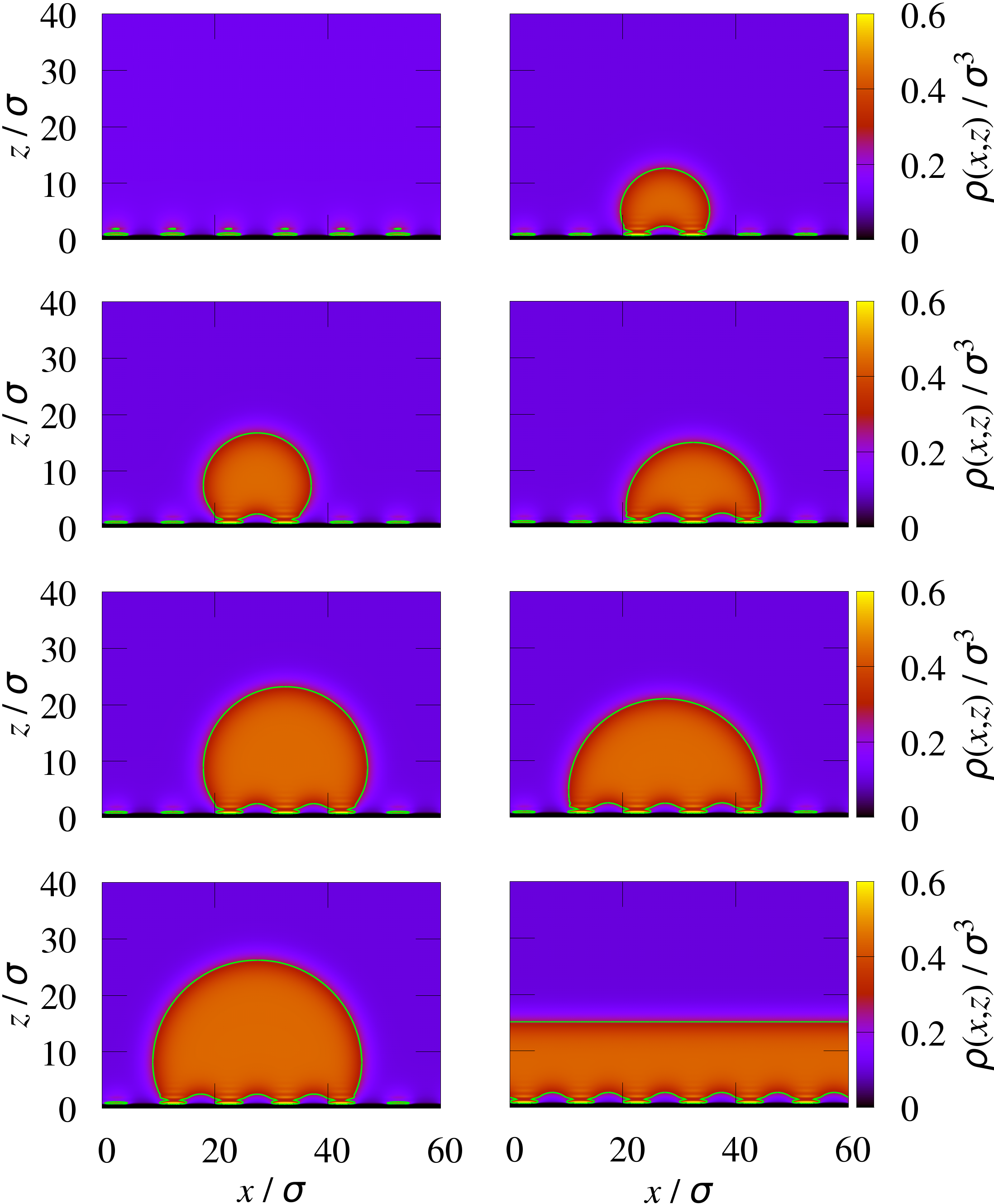}}
 \caption{Two-dimensional density profiles of coexisting liquid morphologies corresponding to states denoted by squares in Fig.~\ref{pd5}a. Contour lines
 corresponding to the arithmetic mean of coexisting bulk liquid and vapour densities, $(\rho_l+\rho_g)/2$, are also shown.}
  \label{dens_prof5}
\end{figure}

To begin, we consider the neutral substrate formed by equal portions occupied by attractive and repulsive stripes. This means that the composition
parameter  $\chi\equiv L_1/L=0.5$, where the width of the attractive stripes $L_1=5\,\sigma$ and the wall periodicity thus $L=10\,\sigma$. In
Fig.~\ref{pd5}a we display the dependence of the free energy per unit length on the fluid particle density. The results  are obtained from DFT by
solving Eq.~(\ref{el}) as a subject to various initial conditions which lead to different, stable or metastable, fluid configurations. The
equilibrium states correspond to the global minimum of the constrained free energy which is denoted by the solid line. The solid line exhibits
several kinks which reflect the presence of  metastable extensions of the free energy (dotted lines). This means that the system undergoes several
first-order phase transitions at which the fluid morphology changes abruptly. For low densities, the stable configurations correspond to microscopic
layers adsorbed at the hydrophilic stripes. The coverage and thus the height of the layers grows gradually with the density and so does the total
surface free energy as the total liquid-gas interface increases. However, this continuous process terminates at the density $\rho_{\rm
av}\approx0.12\sigma^{-3}$ where the system changes its morphology, such that a liquid drop forming a bridge between two neighboring attractive
stripes occurs. In contrast to the previous state, this configuration breaks the system symmetry generated by the wall potential with the location of
the bridging drop being ambiguous, as there is no free-energy cost for shifting the drop by an integer multiple of $L$. This solution is stabilized
by imposing the constraint on the number of the fluid particles, as opposed to a grand-canonical treatment of DFT in which case all the energetically
equivalent states (with a different location of the drop) would be averaged to give either a low density state (separated layers) or a high density
state (liquid film with a corrugated, $L$-periodic interface and gas voids above the repulsive stripes). At the fixed density, the morphological
transition is simply a consequence of lowering the total interfacial area. However, the comparison of the coexisting density profiles as displayed in
Fig.~\ref{dens_prof5} (top row) reveals that the decrease in the surface free energy is \emph{not} due to the bridging itself but because it
decreases the local coverage and thus the interface area \emph{outside} of the drop.

\begin{figure}[]
\centerline{\includegraphics[width=\linewidth]{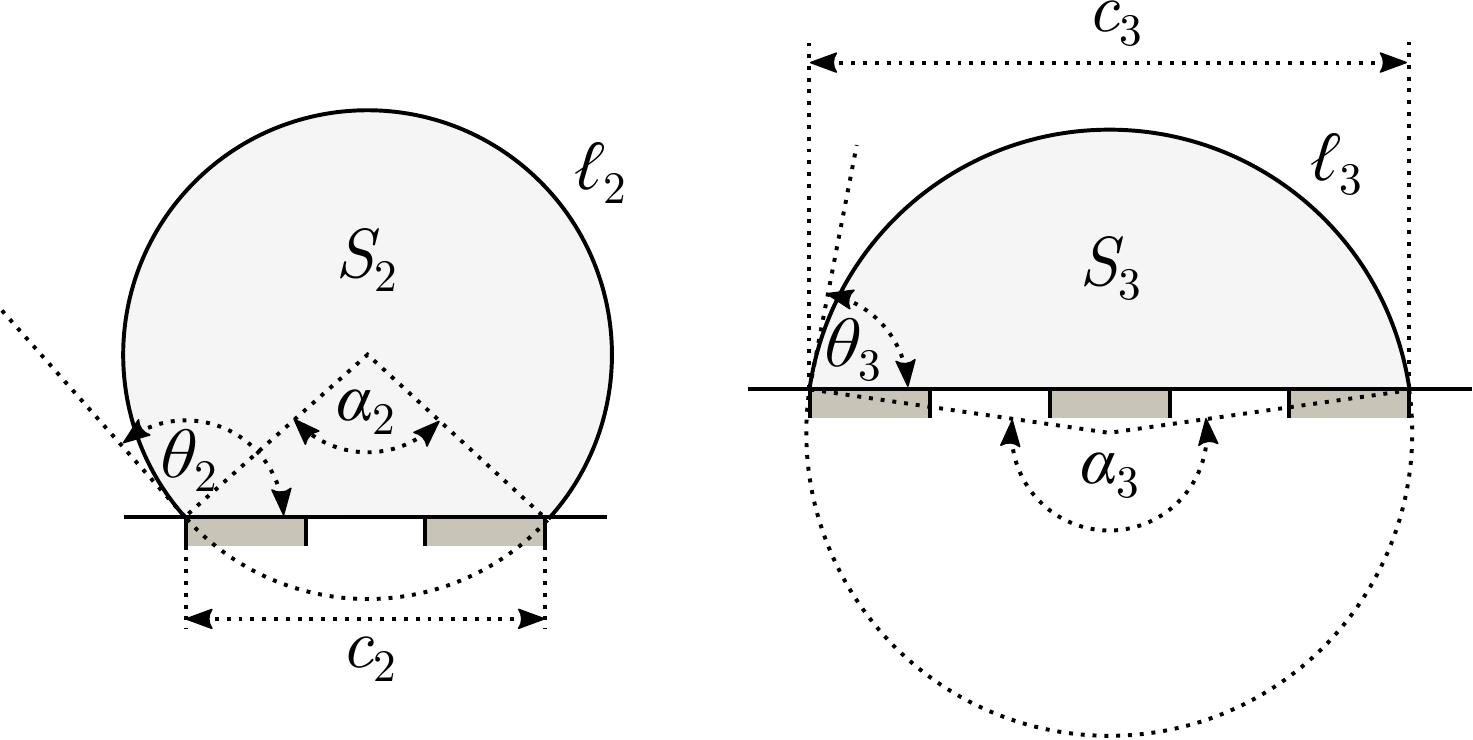}}
 \caption{Sketch of a drop spanning two (left) and three (right) attractive stripes. The
sketch shows the meaning of the contact angles $\theta_i$, the opening angles $\alpha_i$, the drop areas $S_i$, the chords $c_i$ and the contour
lengths $\ell_i$, all used in the macroscopic description (Eqs.~(\ref{delta_F})--(\ref{arc})) of the drop shapes. The subscript $i$ refers to the
number of connected attractive stripes.}
  \label{sketch_cont}
\end{figure}

By increasing the fluid density still more, the system eventually experiences another morphological transition, such that the new state now contains
a  liquid drop spanning three attractive states. This transition is substantially weaker than the previous one, since the change in the fluid
morphology is much less pronounced  than that breaking the system symmetry. Moreover, the coexisting density profiles shown in the second row of
Fig.~\ref{dens_prof5} suggest that also the mechanism of the transition is different and is now dominated by the free-energy change associated with
the bridging, since the coverage outside of the drop does not appreciably change. Therefore, the shapes of the coexisting drops can roughly be
estimated by balancing the corresponding surface free energies per unit length $F_{2}=F_{3}$ and by imposing a complimentary condition fixing the
droplet area $S$ (see Fig.~\ref{sketch_cont}). Hereafter the subscript $i=2,3,\ldots$ refers to a state containing a drop spanning over $i$
attractive stripes. The free energy change $\Delta F=F_3-F_2$ due to the formation of a drop spanning three attractive and bridging over two
repulsive stripes is
 \bb
 \Delta F=\gamma(\ell_3-\ell_2)+\gamma L_2\,.\label{delta_F}
 \ee
Here, the first term is the free-energy change in the outside liquid-gas area, which, per unit length, is given by the decrease in the length
$\ell_i$ of the arc approximated by the circular segment with the chord of $c_i=iL_1+(i-1)L_2$. The second term is the free-energy change due the
covering, by liquid, a repulsive stripe with Young's contact angle $\theta=\pi$, hence the corresponding wall-liquid surface tension satisfies
$\gamma_{wl}=\gamma_{wg}+\gamma$.


If $\alpha_i$ denotes the central angle (in radians) of the drop (see Fig.~\ref{sketch_cont}), the area of the circular segment spanning over $i$
attractive stripes is
 \bb
 S_i=\frac{c_i^2(\alpha_i-\sin\alpha_i)}{8\sin^2\left(\frac{\alpha_i}{2}\right)} \label{area}
 \ee
 and its arc length is
  \bb
 \ell_i=\frac{(2\pi-\alpha_i)c_i}{2\sin\left(\frac{\alpha_i}{2}\right)}\,.  \label{arc}
 \ee
The contours of the drops then can be determined by solving the set of coupled equations $\Delta F=0$ and $S_2=S_{3}$  for $\alpha_2$ and $\alpha_3$
which, in turn, determines the corresponding contact angles $\theta_i=\pi-\alpha_i/2$.

By substituting for the values of the chords $c_2=15\,\sigma$ and $c_3=25\,\sigma$ into Eqs.~(\ref{delta_F})--(\ref{arc}) we find the estimated
values of the contact angles: $\theta_2=124\degree$ and $\theta_3=84\degree$.
This can be compared with our DFT results by analyzing the density profiles. We  define the contour of the drop using a mean density criterion, i.e.
as loci of density profile satisfying $\rho(x,z)=(\rho_g+\rho_l)/2$.
It allows to estimate the contact angles of the drops by constructing a tangent to the contour where the drop meets the wall.  In Fig.~\ref{pd5}b
 we display the DFT dependence of the contact angle on the fluid density.
From the graph, we can read off the contact angles $\theta_2^{\rm DFT}=114\degree$ and $\theta_3^{\rm DFT} =87\degree$ corresponding to the
coexisting drops bridging two and three attractive stripes, respectively, that are in a very reasonable agreement with the previous macroscopic
results.

As the fluid average density is increased still more, the liquid drop grows continuously before another morphological transition occurs, such that
the drop bridges over another repulsive stripe in order to connect four attractive parts. The macroscopic arguments predicting the contact angles of
the coexisting drops can again be tested against the DFT results. Note that the equation expressing the free-energy balance, which is now
$F_{3}=F_{4}$, is completely analogous to (\ref{delta_F}), such that the first term is replaced by $\gamma(\ell_4-\ell_3)$ while the second term,
which accounts for the surface free-energy contribution due to bridging over a repulsive stripe, is unchanged. Complemented with the geometric
equations  (\ref{area}) and (\ref{arc}), one obtains $\theta_3= 117\degree$ and $\theta_4= 90\degree$, which is in an even better agreement with the
DFT results than in the previous case.

Although, upon further increase in density, a morphology involving drop bridging five attractive stripes can still be stabilized, the drop
configuration is already metastable with respect to a thick slab of liquid (with gas-like voids above the repulsive stripes). Therefore, the largest
number of bridged attractive stripes which corresponds to an equilibrium state is four but this number clearly depends on the (lateral) system size;
we will return to this point in the final paragraph of this section. The liquid slab is the ultimate morphology by which the system retrieves its
$L$-periodic symmetry and a further increase in the fluid density leads to a continuous growth in the liquid height.

\subsection{Hydrophobic substrates, $\mathbf{\chi<0.5}$}

\onecolumngrid

\begin{figure}[tb]
\centerline{\includegraphics[width=\linewidth]{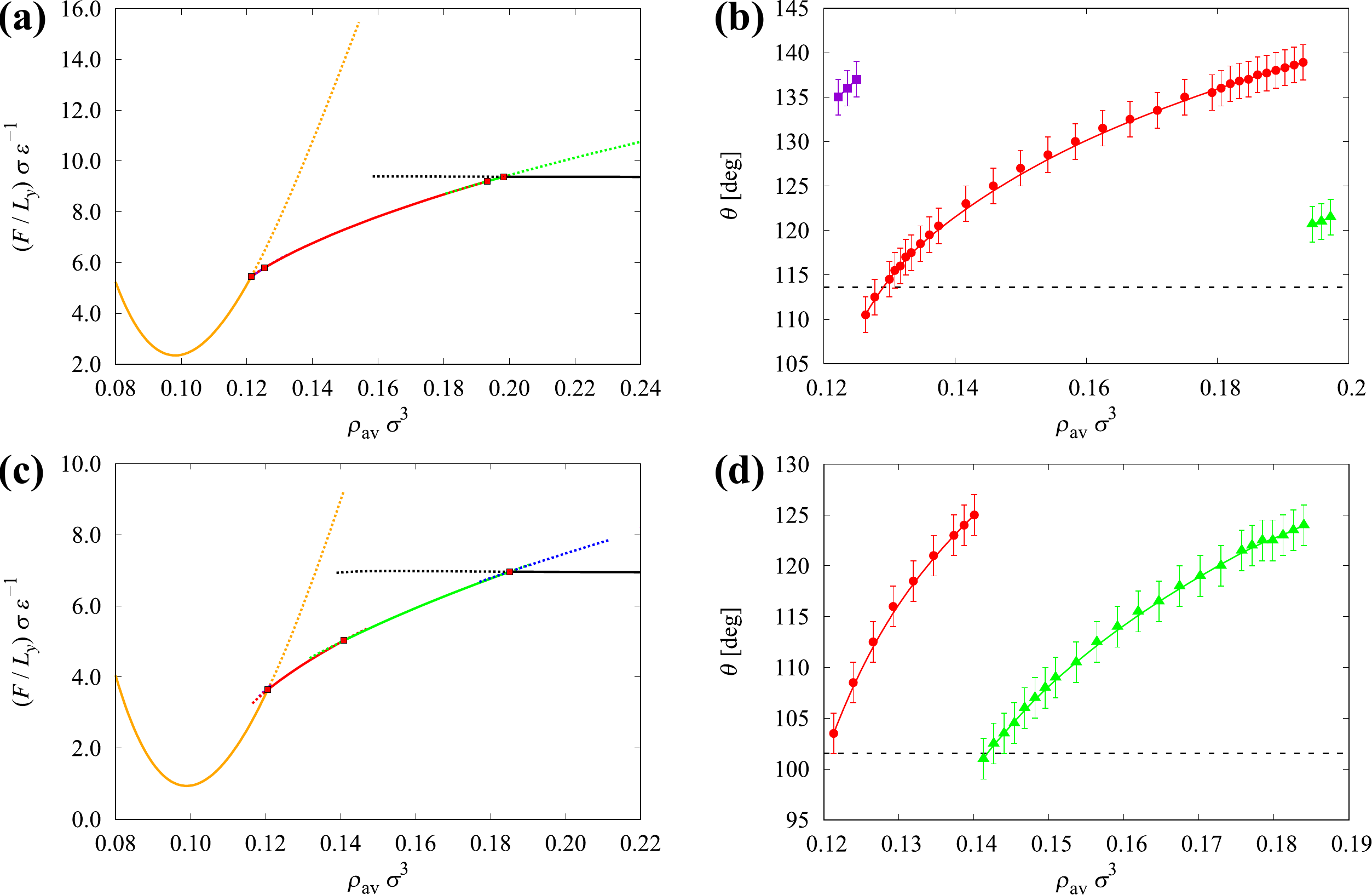}}
 \caption{Left column: Density dependence of the free energy per unit length for the hydrophobic substrates with the composition parameters
a) $\chi=0.3$  and c) $\chi=0.4$. In both cases the stripe widths are $L_1=L_2=5\,\sigma$. The full lines represent the free-energy envelope
corresponding to the equilibrium states of various liquid structures that are separated by first-order morphological transitions highlighted by
filled squares. The configuration corresponding to a microscopic coverage of separated wetting layers (orange curve) is followed by symmetry-breaking
states with a drop attached on one (purple), two (red), three (green) and four (blue) stripes; the free energy branch of a liquid slab morphology is
in black. Also shown are the metastable extensions of the equilibrium states that correspond to local minima of the free energy functional
(\ref{frho}) depicted by dashed lines. Right column: Contact angle density dependence of the droplet configurations for the corresponding substrates
as obtained from the DFT density analysis. The results correspond to stable configurations of a drop attached on one (purple squares) two (red
circles) and three (green upward triangles) attractive stripes. The symbols are connected by lines to guide the eye and vertical lines represent
estimated inaccuracy in the determination of the contact angle. The dashed horizontal line shows the value of the Cassie's apparent contact angle as
given by Eq.~(\ref{cassie}).} \label{pd34}
\end{figure}

\twocolumngrid




We now turn our attention to substrates which are predominantly formed by repulsive, rather than attractive parts and which we thus term as
hydrophobic. Keeping the system periodicity fixed to $L=10\,\sigma$, let us first consider the substrate with the composition parameter $\chi=0.4$
($L_1=4\,\sigma$ and $L_2=6\,\sigma$). Now, in comparison with the previous neutral case, the fluid phase behavior does not allow for the formation
of stable drops spanning four attractive parts, so that the phase diagram shown in Fig.~\ref{pd67}a displays only three morphological transitions.
These are, however, distributed over a larger interval in the fluid density, since the formation of the thick liquid film occurs, as expected, at a
larger density now.
 The macroscopic predictions  can
now be applied for a single transition only involving two and three attractive stripes, with the chords of the circular segments $c_2=14\,\sigma$ and
$c_3=24\,\sigma$. The resulting values of the contact angles are $\theta_2=135\degree$ and $\theta_3=99\degree$ which is again in a good agreement
with those obtained from DFT that are displayed in Fig.~\ref{pd67}b.



\begin{figure}[]
\centerline{\includegraphics[width=\linewidth]{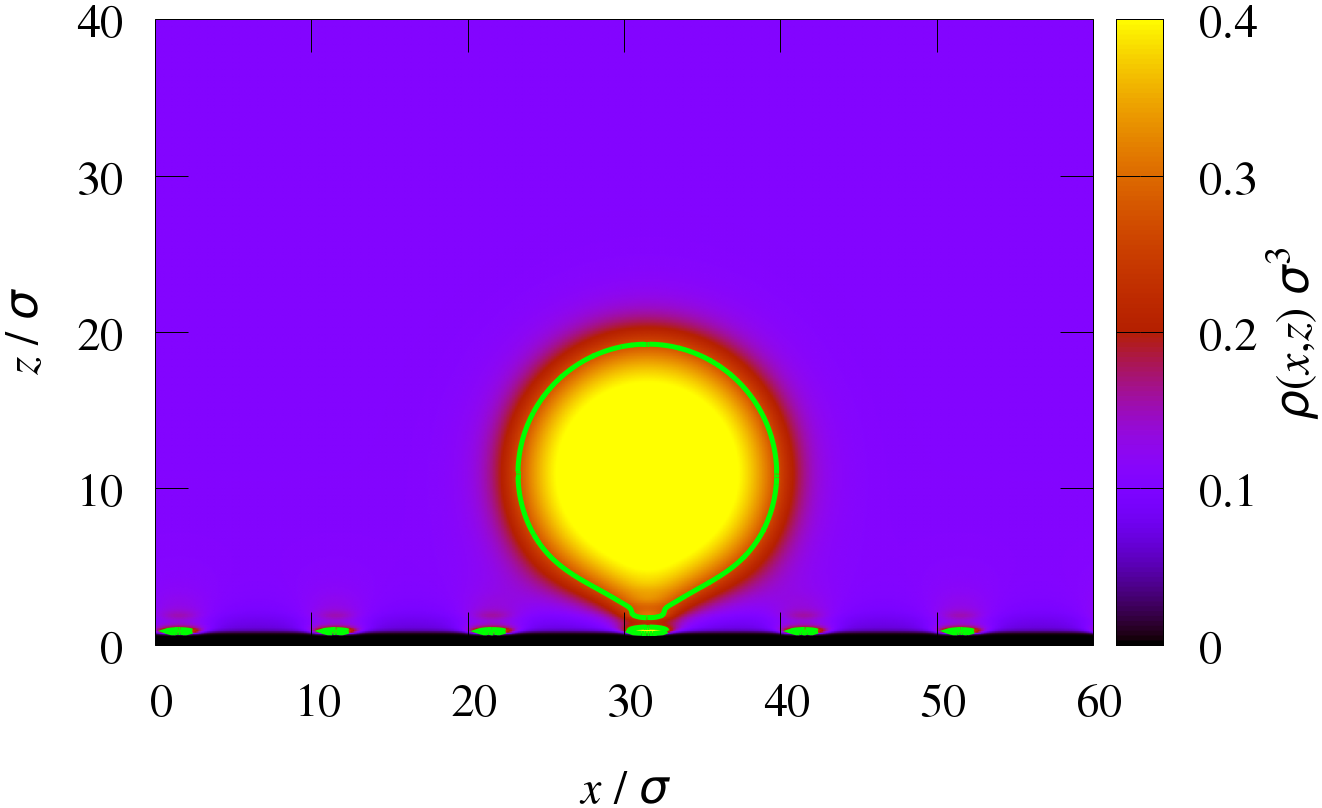}}
 \caption{Density profile of a one-stripe drop for the substrate composition $\chi=0.3$ and the average fluid density $\rho_{\rm
av}=0.125\,\sigma^{-3}$.}  \label{d1}
\end{figure}

Phase behavior of the fluid adsorbed at even more hydrophobic substrate characterized by the composition parameter $\chi=0.3$, is shown in
Fig.~\ref{pd34}c . The interval in the fluid density within which the morphological transitions occur is still larger and involves four first-order
transitions points again, as in the case of the neutral substrate, but over a different spectrum of morphologies. Interestingly and perhaps somewhat
counterintuitively, a morphology of a drop attached to a single attractive stripe is now possible, even though the width of the stripe is as small as
$L_1=3\,\sigma$. This configuration, shown in Fig.~\ref{d1}, is more stable at low densities than the one corresponding to a drop spanning two
attractive stripes in view of a large free-energy barrier to be overcome to form a bridge over a long repulsive bit. For the same reasons, the
configuration involving a drop spanning more than three stripes are now absent and not even metastable. The contact angle density dependence for the
drop morphologies is shown in Fig~\ref{pd34}d. The values of the contact angles obtained from the macroscopic description for the coexisting
one-stripe and two-stripe drops are $\theta_1=163\degree$ and $\theta_2=100\degree$, while for the coexistence between two-stripe and three-stripe
drops the predictions are $\theta_2=145\degree$ and $\theta_3=115\degree$.

\begin{figure}[]
\centerline{\includegraphics[width=\linewidth]{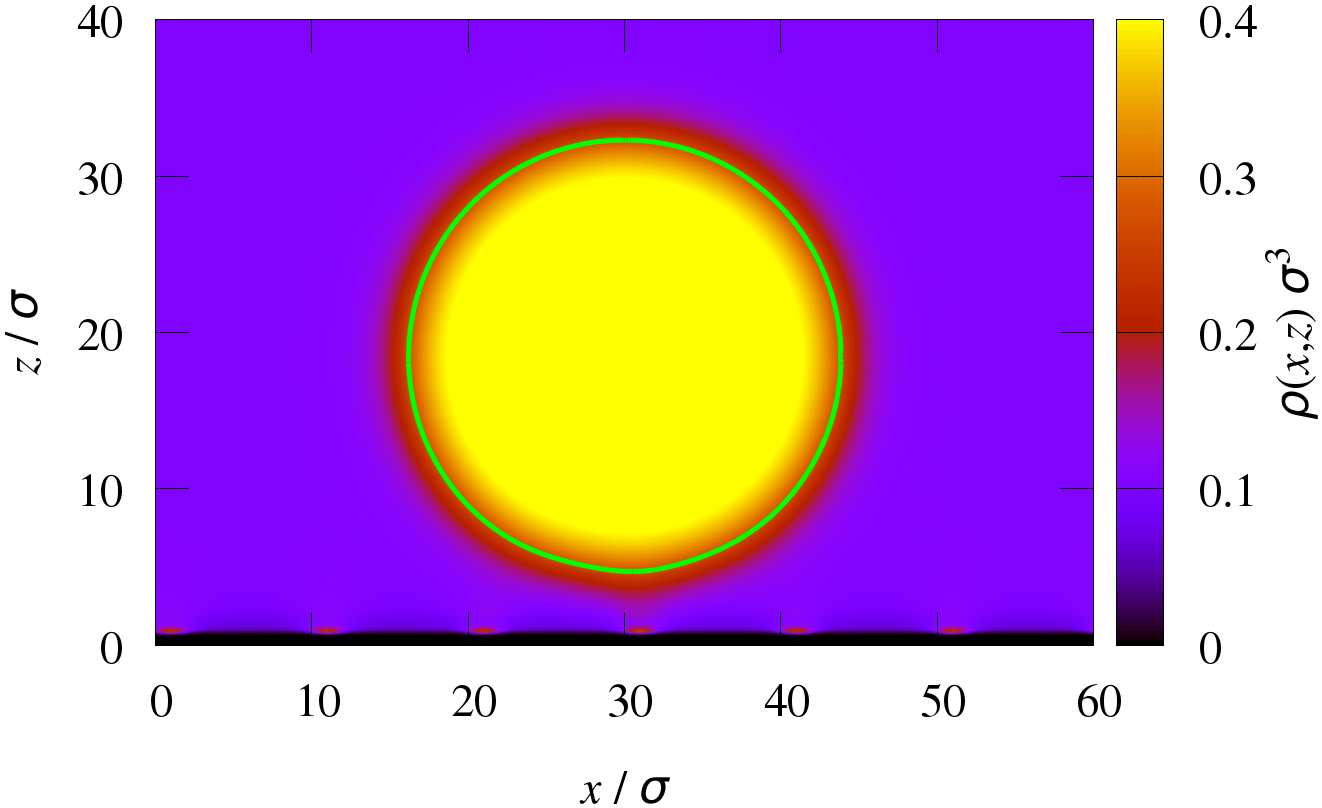}}
 \caption{Density profile of a non-sessile drop for the substrate composition $\chi=0.2$ and the average fluid density $\rho_{\rm av}=0.158\,\sigma^{-3}$.}
  \label{float}
\end{figure}

However, For strongly hydrophobic substrates with the composition parameter $\chi\le0.2$ the situation becomes quite different. In these cases, the
effective contact angle is already so high that the substrates essentially tend to avoid any contact with the liquid and act effectively as a hard
wall. Therefore, sessile drops are not stabilized anymore and the systems experience only one morphological transition corresponding to a formation
of a single drop floating \emph{above} the wall as the density is sufficiently high, see Fig.~\ref{float}. The location of the drop is again largely
arbitrary and the averaging over all droplet morphologies would lead to a complete drying scenario with a gas film intruding at the wall-liquid
interface.

\subsection{Hydrophilic substrates, $\mathbf{\chi>0.5}$}

\onecolumngrid

\begin{figure}[tb]
\centerline{\includegraphics[width=\linewidth]{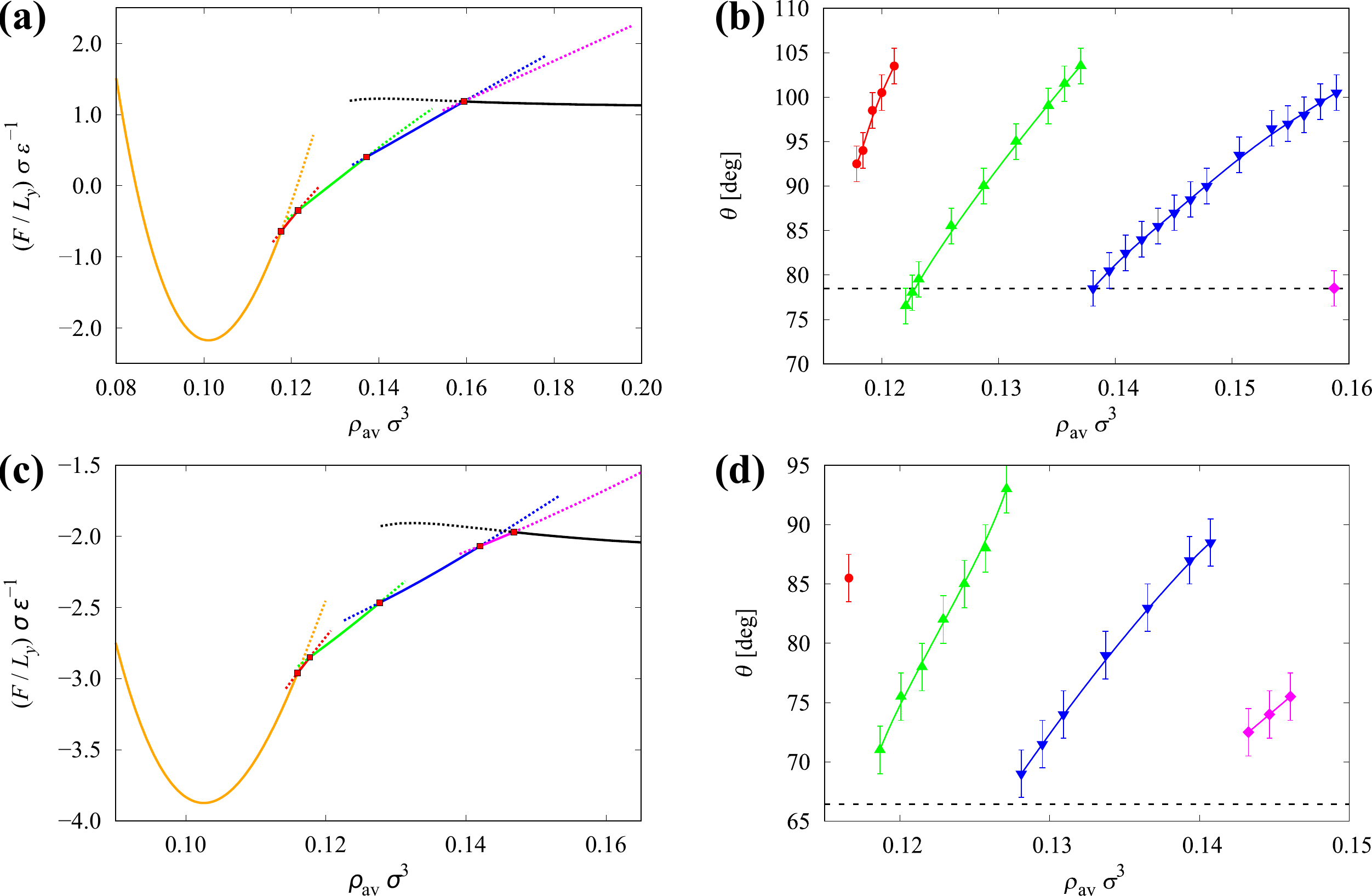}}
 \caption{Left column: Density dependence of the free energy per unit length for the hydrophilic substrates with the composition parameters
a) $\chi=0.6$  and c) $\chi=0.7$. In both cases the stripe widths are $L_1=L_2=5\,\sigma$. The full lines represent the free-energy envelope
corresponding to the equilibrium states of various liquid structures that are separated by first-order morphological transitions highlighted by
filled squares. The configuration corresponding to a microscopic coverage of separated wetting layers (orange curve) is followed by symmetry-breaking
states with a drop attached on two (red), three (green), four (blue) and five (magenta) stripes; the free energy branch of a liquid slab morphology
is in black. Also shown are the metastable extensions of the equilibrium states that correspond to local minima of the free energy functional
(\ref{frho}) depicted by dashed lines. Right column: Contact angle density dependence of the droplet configurations for the corresponding substrates
as obtained from the DFT density analysis. The results correspond to stable configurations of a drop attached on  two (red circles), three (green
upward triangles), four (blue downward triangles) and five (magenta diamonds)attractive stripes. The symbols are connected by lines to guide the eye
and vertical lines represent estimated inaccuracy in the determination of the contact angle. The dashed horizontal line shows the value of the
Cassie's apparent contact angle as given by Eq.~(\ref{cassie}).} \label{pd67}
\end{figure}

\twocolumngrid



Let us now turn our attention to substrates where the dominating portion of the surface is occupied by attractive stripes and for which the
application of Cassie's law leads to an effective contact angle $\theta^*<\pi/2$. Adsorption properties of only slightly hydrophilic substrate with
the composition parameter $\chi=0.6$ can be distinguished from those at the neutral substrate in two aspects, as can be observed from the phase
diagram shown in Fig.~\ref{pd67}a. Firstly, the density interval within which the morphological transitions occur is markedly shorter which is
because the transition to the complete wetting state (liquid slab) can be accomplished more easily now and thus occurs at the lower density. Secondly
and more importantly, the system now exhibits a triple point where four-stripes drop, five-stripes drops and liquid film morphologies all coexist.
Note also, that except for the triple point itself, the five-stripes drop morphology is always metastable, although its extension is substantially
larger than in the neutral case. Interestingly, the predicted values of the contact angles for coexisting states based on
Eqs.~(\ref{delta_F})--(\ref{arc}): $\theta_2=115\degree$ and $\theta_3=73\degree$, $\theta_3=108\degree$ and $\theta_4=79\degree$,
$\theta_4=104\degree$ and $\theta_5=83\degree$ are even in a better agreement than in the neutral case, as can be seen from the comparison with the
DFT results shown in Fig.~\ref{pd67}b.



For moderately hydrophilic substrate with $\chi=0.7$, the morphology involving a five-stripes drop becomes stabilized before the liquid slab
configuration forms, see Fig.~\ref{pd67}c, and it is the first case when the system undergoes five morphological transitions. Moreover, the
transitions are now packed over a still shorter period of density; this is also reflected by the contact angle behaviour which is now much more
sensitive to the change in density than in the previous cases and forms a sequence of nearly straight lines rather than concave curves as depicted in
Fig.~\ref{pd67}d. Again, the values of the contact angles for the coexisting states as given by Eqs.~(\ref{delta_F})--(\ref{arc}):
$\theta_2=104\degree$ and $\theta_3=62\degree$, $\theta_3=97\degree$ and $\theta_4=68\degree$, $\theta_4=93\degree$ and $\theta_5=71\degree$, are in
a very good agreement with the DFT results, except for the smallest, two-stripe drop. Qualitatively same behavior is observed also for $\chi=0.8$.

However, for strongly  hydrophilic substrates the adsorption scenario changes dramatically. We find that there exists a critical value of the
composition parameter $\chi_c\sim0.85$ such that the systems with $\chi>\chi_c$ do not exhibit any symmetry breaking and the adsorption is a fully
continuous process. Here, the repulsive stripes are already so narrow that the bridging barrier vanishes, so that the microscopic wetting layers get
connected directly to form a liquid film in a continuous manner. Hence, this class of substrates act effectively  as the homogeneous wall with
$\chi=1$.

\begin{figure}[]
\centerline{\includegraphics[width=\linewidth]{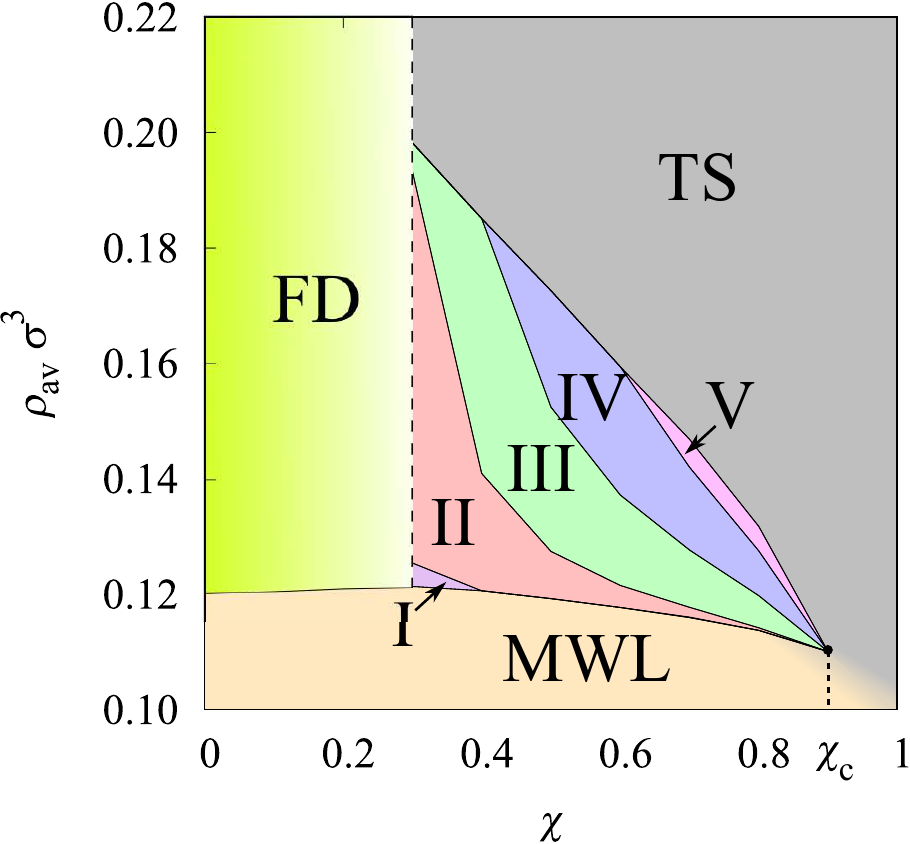}}
 \caption{The global phase diagram showing the equilibrium fluid morphologies for different substrate composition $\chi$ and average fluid density $\rho_{\rm av}$.
The inscriptions represent the following morphologies: microscopic wetting layers (MWL), thick slab (TS), floating drop (FD) and the Roman numerals
refer to a bridging drop spanning the corresponding number of attractive stripes. The critical composition $\chi_c$ above which the system fluid
adsorption is continuous is also displayed.}
  \label{pdg}
\end{figure}

In conjunction  with the results shown in the previous paragraph, we summarize the main results by constructing the global phase diagram which
depicts the most stable configurations for the given values of $\chi$ and $\rho_{\rm av}$, see Fig.~\ref{pdg}. For our model consisting of six
attractive and six repulsive stripes, and with a wavelength of $L=10\,\sigma$, we could stabilize eight different configurations. Only two of them,
corresponding to a low-density coverage and a thick liquid film, follow the symmetry of the wall potential. The remaining six configurations involve
a liquid drop which necessarily breaks the symmetry of the system. Typically, the liquid drop is sessile, attached at one or more attractive stripes
and bridging over the inner (if any) repulsive parts. For strongly hydrophobic walls with $\chi<0.3$, though, the liquid drop unbinds from the wall
and floats freely in the gas phase. All the morphological changes are of first-order except for the transition from separate microscopic wetting
layers to a thick liquid film configurations, occurring at high values of $\chi$, which is continuous.

\subsection{Impact of periodicity and system size}

\begin{figure}[]
\centerline{\includegraphics[width=\linewidth]{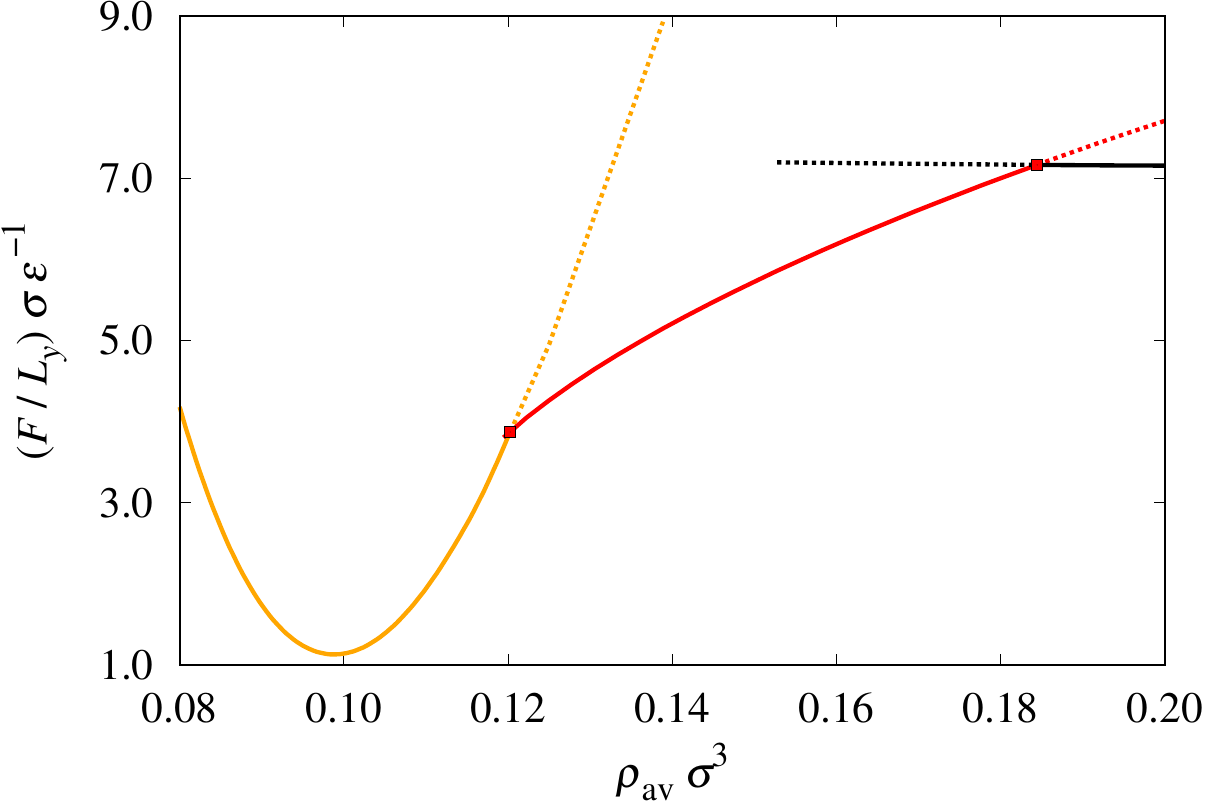}}
 \caption{Free energy density dependence for the neutral substrate ($\chi=0.5$) and periodicity $L=4\,\sigma$. Apart from the low coverage and
 the thick liquid film branches, the system develops continuously over drop-like structures with a gradually increasing number of attractive stripes
 onto which the drop is attached.}
  \label{pd22}
\end{figure}

We further investigate the influence of the substrate periodicity $L$ and the size of the system on the fluid phase behavior. We first focus on the
neutral substrates, $\chi=0.5$, with the same system size of $L_x=60\,\sigma$ as before, but with a finer structure, i.e. with a shorter periodicity.
In particular, for $L=6\,\sigma$ we have still observed a sequence of first-order transitions between morphologies including drops spanning over up
to six attractive stripes but the transitions become extremely weak and the metastable extensions of the drop configurations hardly detectable. As
the periodicity is still lowered, however, we already could not detect any multiple solutions and the low coverage and the thick liquid film
configurations are connected via a series of continuously developing symmetry-breaking morphologies with a drop spanning gradually an increasing
number of attractive stripes. This is illustrated in Fig.~\ref{pd22} showing the free energy density dependence for $L=4\,\sigma$ which now consists
of only three branches.

\begin{figure}[]
\centerline{\includegraphics[width=\linewidth]{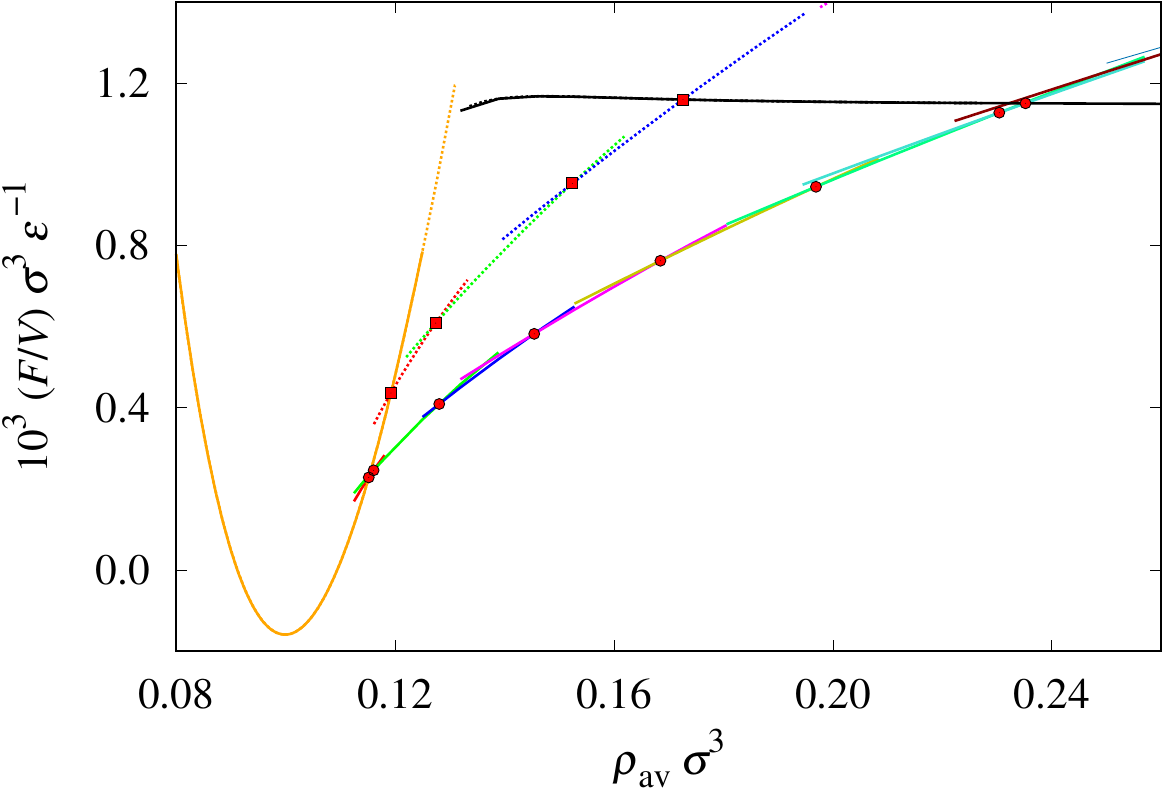}}
 \caption{The comparison of the free energy (per unit volume) density dependence  of two differently wide systems: $L_x=60\,\sigma$ (dotted lines) and
 $L_x=120\,\sigma$ (full lines). In both cases, the substrates are neutral ($\chi=0.5$) and of a periodicity $L=10\,\sigma$.    }
  \label{comp_5_5}
\end{figure}

As a final point, we want to discuss the impact of the system width $L_x$. If this dimension is increased by considering more stripes of the wall,
one expects that the number of symmetry-breaking structures also increases simply because the liquid drop can adopt a larger spectrum of states that
differ by the number of bridged attractive stripes. In order to get a more detailed insight into the finite-size effects, we have performed DFT
calculations for the composition parameter $\chi=0.5$ and periodicity of $L=10\,\sigma$ but with  $L_x=120\,\sigma$, i.e. by considering a system
consisting of twelve attractive (and repulsive) stripes. The height of the system is unchanged, $L_z=60\,\sigma$. The comparison of the phase
behaviour between the large ($L_x=120\,\sigma$) and the small ($L_x=60\,\sigma$) systems is shown in Fig.~\ref{comp_5_5} where we now display the
density dependence of the free-energy density $F/(L_xL_yL_z)$. We can see that the parts of the free energies corresponding to the lowest and highest
densities, i.e. morphologies that do not break the system symmetry, essentially overlap which is because the free-energy for these configurations is
extensive. This, however, does not apply for the remaining symmetry-breaking morphologies containing a single drop, in which case the free energy
does not scale with $L_x$ anymore. Here, the free-energy density for the larger system is shifted to lower values and the density differences between
the successive transitions are reduced, except, of course, those that are not present in the smaller system; these include (stable) drops that
connect six to eight attractive stripes.

Now, the natural question to pose is: what is the phase behaviour of an infinitely wide system $L_x\to\infty$? There are several conceivable
scenarios: i) the system exhibits an infinite number of first-order morphological transitions; ii) there is a finite number of morphological
first-order transitions that eventually become continuous; iii) in the limit of $L_x\to\infty$ all the transitions collapse into a single point; iv)
the adsorption proces becomes continuous.

\begin{figure}[]
\centerline{\includegraphics[width=\linewidth]{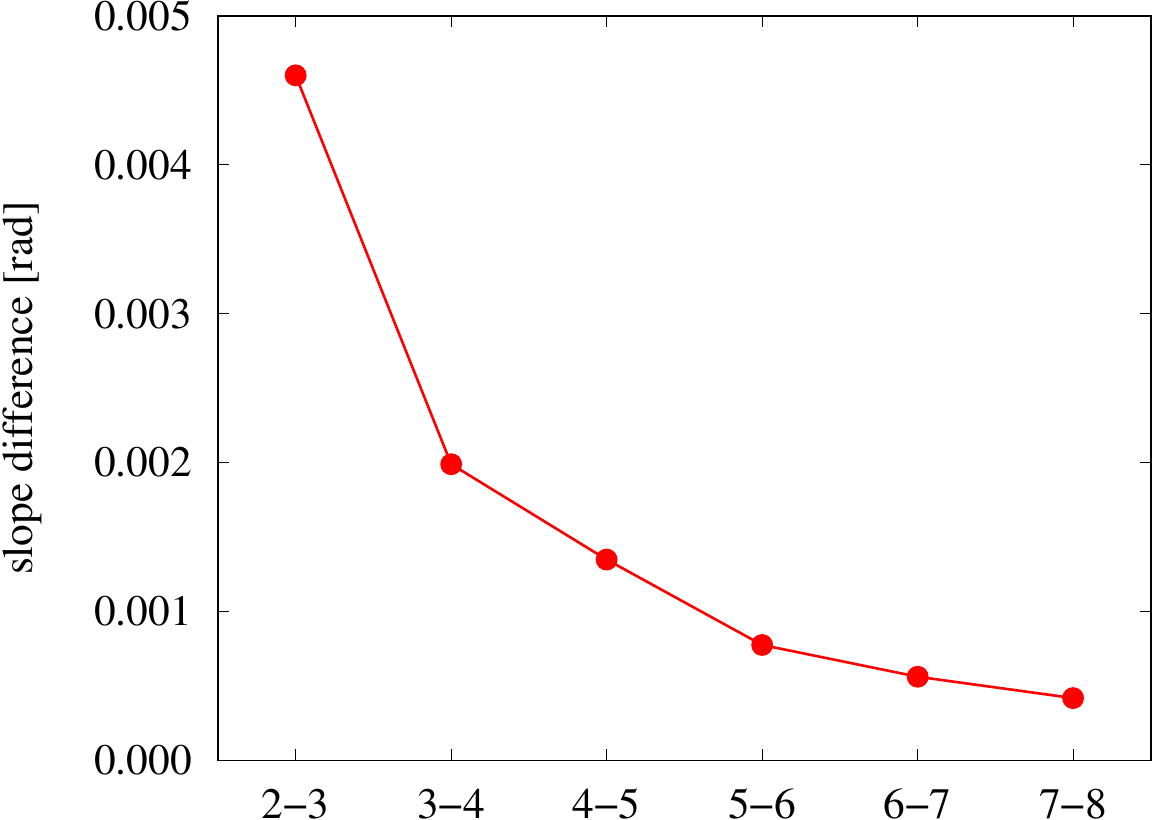}}
 \caption{Slope differences of free-energy branches associated with the change of liquid morphology for the neutral substrate of the width
 of $L_x=120\,\sigma$. The horizontal axis denotes the number of attractive stripes which the bridging drop connects.}
  \label{ptr}
\end{figure}

The scenario iv) would only apply for a very fine structure of the wall with $L\lesssim 4\,\sigma$, or for a very high value of the composition
parameter $\chi>\chi_c$, as discussed previously. Although the density intervals separating the morphological transitions are reduced with the size
of the system, the presence of sufficiently wide repulsive stripes still generates a free-energy barrier which must be overcome to connect wetting
layers on the attractive stripes. It is still admissible, though, that the system in the limit of $L_x\to\infty$ exhibits a single first-order
transition from the low- to the high-coverage state and, indeed, this is what happens in an open system \cite{pos2019}. However, this scenario is
inconsistent with our DFT results shown in Fig.~\ref{comp_5_5}, since the non-extensive free-energy branch clearly does not disappear in the limit of
$L_x\to\infty$ and there is no reason to expect that the transitions corresponding to a formation of the drop becomes rounded in this limit.
Therefore, we are left only with the first two options. Although it is unreasonable to expect that there exists a singular point in the system
associated with the finite critical number of attractive stripes $n_c$, such that the bridging transitions loose their first-order character when
$n_c$ stripes are involved, it is clear that the ``strength'' of the transitions must decay sufficiently rapidly with the size of the bridging drop,
since each transition is accompanied with the change in the free-energy slope, the sum of which must remain finite for $L_x\to\infty$. This is indeed
supported by the results displayed in Fig.~\ref{ptr} showing the slope difference between consecutive free-energy branches determined at the
morphological transitions, as obtained from the DFT calculations for the large system ($L_x=120\,\sigma$). As expected, the slope difference decays
monotonically with the number of bridged attractive stripes $n$ and any reasonable extrapolation of this dependence suggests that the asymptotic
value of the graph is zero, such that the slope difference is of the order of $10^{-5}\,{\rm rad}$, i.e. practically indistinguishable, already for
$n\approx20$.


\section{Summary and concluding remarks}

In this work, we studied adsorption properties of molecularly patterned walls using a non-local density functional theory and simple macroscopic
arguments. We have considered planar model substrates formed by regularly alternating stripes of contrasting wettability. To this end, the potential
of one type of the stripes was set to be strongly attractive, such that it induces complete wetting (Young's contact angle $\theta=0$), while the
other type of the wall is purely repulsive and tends to be complete dry ($\theta=\pi$).

We assumed that the stripes are macroscopically  long but the remaining dimensions of the system are microscopic (i.e., of the order of molecular
diameters) and that they confine a fixed amount of fluid particles. This allows for a formation of fluid structures that would not be stable in open
systems, in which case the fluid structure must follow the symmetry of an external field exerted by the wall. In this way, we were able to observe
various liquid morphologies typically separated by first-order transitions, as the number density of the fluid atoms varies. We have studied the
mechanism leading to complete wetting (or drying) of the wall depending on the composition parameter $\chi$ (characterizing the relative width of the
attractive and repulsive stripes), the system wavelength and the system size.

The main results are summarized in the diagram shown in Fig.~\ref{pdg}. At low densities, the wall is generally covered by a microscopic amount of
liquid concentrated at the attractive stripes.  Except for the extreme cases of the wall composition, the increase in the fluid density leads to a
formation of a thick slab of liquid  via a sequence of fluid morphologies breaking the system symmetry; these correspond to a liquid drop bridging
several attractive stripes. The drop size increases continuously with density via increasing its contact angle before it adopts a more favourable
formation by bridging another wetting layer at the neighbouring attractive stripe. This morphological change which changes the shape of the drop
abruptly is typically accompanied by a weak first-order transition.

However, there are several important exceptions regarding the nature of the morphological transitions. The transitions between different drop
configurations all become rounded  if the wavelength of the system $L$ is below $L_c$, the magnitude of which is a few molecular diameters. Moreover,
there exists a critical wall composition which for our model is $\chi_c\approx0.85$, such that the entire adsorption process is continuous for
$\chi>\chi_c$, akin to complete wetting on a homogeneous wall. In both cases, the character of the rounding of the transitions below the critical
values is analogous to the presence of the critical radius for bridging of spherical nanoparticles \cite{bridging1,bridging2}. In the opposite
extreme case of a strongly hydrophobic wall with $\chi\lessapprox0.2$ the system exhibits a single morphological transition from a microscopic
coverage state to a state containing an unbind liquid drop floating above the wall.

Finally, we have argued that in the thermodynamic limit, i.e. when the lateral size of the system is macroscopic (infinite), the wall (of a moderate
value of $\chi$ and for $L>L_c$) becomes completely wet via a sequence of morphological first-order transitions that are, however, gradually weaker
as the size of the bridging drop increases, such that they eventually become effectively continuous. This mechanism sharply contrasts  with that for
an open system, in which case the system may only undergo a single, pre-wetting transition \cite{pos2019}.

We have also formulated a simple macroscopic theory to predict the contact angles of coexisting bridging drops, i.e. the contact angle of a drop
before and after morphological transition when it forms another bridge. Here we assumed that the drop adopts a shape of a circular segment with an
appropriate chord corresponding to a number of bridged stripes and that the volume of the drop is conserved at the transition. The comparison with
the results obtained from an analysis of the DFT density profiles revealed a surprisingly good agreement especially for drops spanning larger number
(three or more) of bridged stripes.

We conclude with two remarks pertinent to the relevance of our findings:

Firstly, in our model, the ``hydrophobic'' parts of the wall were considered to be purely repulsive. Clearly, this is a simplification of any
realistic situation, since attractive forces are ubiquitous in nature. However, the absence of the attraction and also the implied fact that Young's
contact angle of these stripes is $\theta=\pi$  are not crucial for our model and our conclusions would not qualitatively change if the
``hydrophobic'' stripes were partially wet (instead of completely dry); if it was the case (which would require another interaction parameter) the
bridging of microscopic wetting layers into a drop would still need to overcome a free-energy barrier, although smaller than in the present case. We
thus expect a shift in the critical values of $\chi_c$ (downwards) and $L_c$ (upwards).

Secondly, our DFT analysis is of a mean-field character and largely neglects the interfacial fluctuations which may destroy some of the phase
transitions. Indeed, any morphological phase transition  involving a cylindrically shaped drop will strictly speaking be rounded due to fluctuations
along the stripes; these will break the cylindrical drops up into liquid and gas domains in view of their pseudo one-dimensional character, even if
the substrate width is infinite \cite{privman1983}.  However, the rounding is proportional to the Boltzmann factor $\exp[-\beta\gamma S]$ where
according to our macroscopic result $S\propto (nL-L_2)^{2}$ is the area of the lateral cross-section of droplets covering $n$ hydrophilic stripes.
Therefore, albeit rounded, the transition is expected to be very sharp out of the critical region (where $\gamma$ is very small), especially for
larger drops. On the other hand, in contrast to free cylindrical liquid streams, the cylindrical drops should be resistent towards the
Plateau--Rayleigh instability \cite{plateau, rayleigh} since they are pinned to the edges of the hydrophilic stripes which does not thus allow for
pinching of the cylinder into drops. However, it is conceivable that for substrates with a small difference in the surface interaction exerted by the
hydrophilic and hydrophobic stripes, in which case the pinning is much weaker, the Plateau--Rayleigh instability can also be relevant. Finally, the
effect of the capillary-wave fluctuations which is characterized by the roughness parameter\cite{schick} $\xi_\perp\propto\sqrt{\ln(nL-L_2)}$ (in
units of $\sigma$) is completely negligible for the present model.

\begin{acknowledgments}
\noindent This work was funded by the Czech Science Foundation, Project No. GA17-25100S. M.P. acknowledges the financial support from specific
university research (MSMT No 21-SVV/2019).
\end{acknowledgments}

\bibliographystyle{aip}
\bibliography{reserse}

\end{document}